\documentclass[pre,showpacs,showkeys,preprintnumbers,amsmath,amssymb,superscriptaddress,twocolumn]{revtex4-1}
\usepackage[english]{babel}
\usepackage[T1]{fontenc}
\usepackage[colorlinks=true, pdfstartview=FitV, linkcolor=blue, citecolor=red, urlcolor=magenta, breaklinks=true]{hyperref}
\usepackage[all]{xy}
\usepackage{float} 
\usepackage{amsfonts} 
\usepackage{bm}
\usepackage{graphicx}
\usepackage[all]{xy}
\usepackage{amsmath}
\usepackage{amssymb}
\usepackage{color}
\usepackage{subfigure}
\usepackage{epstopdf}
\usepackage{textcomp}

\def\ben{\begin{eqnarray}}
\def\een{\end{eqnarray}}
\def\be{\begin{equation}}
\def\ee{\end{equation}}

\def\e{\text{e}}

\begin{document}
\title{Evaluating Bohm's quantum force in the scattering process by a classical potential}

\author{W. S. Santana}\email{wanisson.santana@ufob.edu.br}\author{C. Cruz}\email{clebson.cruz@ufob.edu.br}
\affiliation{Grupo de Informa\c{c}\~{a}o Qu\^{a}ntica e F\'{i}sica Estat\'{i}stica, Centro das Ci\^{e}ncias Exatas e das Tecnologias, Universidade Federal do Oeste da Bahia. Rua Bertioga, 892, Morada Nobre I, 47810-059 Barreiras, Bahia, Brazil.}
\author{E. Lima}\email{elisama.lima@ifba.edu.br}
\affiliation{Instituto Federal da Bahia. Rua Gileno de S\'{a} Oliveira, 271, Recanto dos P\'{a}ssaros, 47808-006, Barreiras, Bahia, Brazil.}
\author{F. V. Prudente}\email{prudente@ufba.br}
\affiliation{Instituto de F\'{i}sica, Universidade Federal da Bahia, Rua Bar\~{a}o de Jeremoabo, s/n, Campus de Ondina, 40170-115, Salvador, Bahia, Brazil.}


\begin{abstract}
{In this work, we show an application of the de Broglie-Bohm Quantum Theory of Motion (QTM) as a powerful tool for evaluating Bohm's quantum force in the scattering process of a Gaussian wavepacket by a classical Eckart potential. Our results show that in the absence of a classical potential, the system experiences quantum effects arising from an effective force, intrinsically related to the existence of the wavepacket itself. In contrast, in the scattering by the classical potential, it experiences a quantum force effect even in the absence of any classical force, reinforcing the fact that potentials can act without classical force fields. Thus, this application could be useful to introduce QTM, through the discussion of the concept of Bohm's quantum force, as a classroom working tool instead of merely an alternative interpretation of the quantum theory.}

\end{abstract}
\maketitle

\section{Introduction}

For a long time, the scientific community tried to preserve the classical determinism for quantum events, one of the most relevant and best-structured theories about this theme comes up from the de Broglie-Bohm formulation of the quantum mechanics \cite{HOLLIVRO:93,bricmont2016broglie,sanz2019bohm,styer2002nine,belinsky2019david,bohm1952suggested,bohm85suggested,de1928nouvelle,bacciagaluppi2009quantum}. Based on the conceptions of the pilot wave of de Broglie \cite{de1928nouvelle,bacciagaluppi2009quantum}, David Bohm proposes a theoretical formulation for the quantum mechanics \cite{HOLLIVRO:93,bricmont2016broglie,sanz2019bohm,styer2002nine,belinsky2019david}, in which the quantum events are driven according to an essentially quantum potential, arising from the interaction between the particle and its wave-guide, being responsible for the quantum nature of the events during the system dynamics 
\cite{sanz2019bohm,pinto2013quantum,hasan2016effect,lentrodt2020ab,gonzalez2007quantum,gonzalez2009bohmian,gonzalez2004bohmian,gonzalez2009bohmian2,gonzalez2008effective}.
 
{Recent works has been proposed alternative applications to this formulation \cite{bricmont2016broglie,sanz2019bohm,pinto2013quantum,becker2019asymmetry,sanz2015investigating,batelaan2015dispersionless}, unraveling interesting properties and interpretations for the dynamics of quantum systems \cite{becker2019asymmetry,batelaan2015dispersionless}. The existence of the quantum potential provides one path towards the understanding these in a Newtonian-like view, through the existence of the so-called Bohm's quantum force \cite{maddox2003estimating,becker2019asymmetry,batelaan2015dispersionless}. }  
Recently, Becker \textit{et al.} \cite{becker2019asymmetry}  observed the quantum force predicted by Shelankov \cite{shelankov1998magnetic}, Berry \cite{berry1999aharonov} and Keating \cite{keating2001force} for an Aharonov-Bohm physical system, providing the experimental support for the evidence of the quantum force in the Aharonov-Bohm effect \cite{aharonov1959significance}. 

{In this context, we show in this paper a application of the de Broglie-Bohm Quantum Theory of Motion (QTM) to estimate Bohm's quantum force in the quantum dynamics of a Gaussian wavepacket, with and without the presence of a classical potential . For that, we consider two situations, the first one associated with the free particle case \cite{livre,belinfante1974survey}, and the second one related to a system subjected to the Eckart potential model \cite{gauss,dewdney1982quantum}. The dynamic variables were analyzed through the temporal propagation technique, using the popular and easy to implement finite-difference method, facilitating the reproduction of this analysis for most undergraduate and graduate quantum mechanics students \cite{finite}.}  

Our results show that in the absence of a classical potential, the system experiences quantum effects arising from an effective force intrinsically related to the existence of the wavepacket itself, while the classical determinism is preserved in some way. Moreover, in the scattering by a classical potential, the wavepacket experiences a quantum force effect which depends on the presence of the potential, even in the absence of any classical force field, perceiving it even before the explicit interaction, strengthen the fact that classical potentials can act without force fields and giving us indications that the nature of the Aharonov-Bohm effect can be observed in different classical potentials. 

{Therefore, this application could be used as an introduction to the concept of Bohm's quantum force, presenting the QTM as a useful working tool for study the quantum dynamics, instead of merely an alternative interpretation of the quantum theory.}

\section{de Broglie-Bohm Interpretation}\label{sec-1}

The de Broglie-Bohm QTM presents an interesting interpretation for quantum mechanics, in which the quantum system can be interpreted as two intrinsic counterparts: a wave and a point particle ~\cite{HOLLIVRO:93,maddox2003estimating,sanz2019bohm}. In this context, an individual system comprises one wave, that propagates into spacetime driving the motion of a punctual particle. The wave is mathematically described by a function $\Psi(q_i;t)$, which is a solution of the  Schr\"odinger's equation, in such a way that
\begin{eqnarray}
\Psi(q_i;t)=R(q_i;t)\, \e^{i S(q_i;t)/\hbar}~,
\label{eq:01}
\end{eqnarray}
where  $R=R(q_{i},t)$ and $S=S(q_{i},t)$ are real functions given by:  
\begin{eqnarray}
R(q_{i},t)&=&|\Psi(q_{i},t)| \geq 0, \qquad \forall \qquad  \lbrace q_{i},t\rbrace~, \qquad 
\label{eq:02} \\
\frac{S(q_{i},t)}{\hbar}&=&\tan^{-1}\left(\frac{\mbox{Im}\lbrace\Psi(q_{i},t)\rbrace}{\mbox{Re}\lbrace\Psi(q_{i},t)\rbrace}\right)~.
\label{eq:03}
\end{eqnarray}

Here $S$ can be seen as an action having dimension of $\hbar$.

Considering the functional form of $\Psi(q_i;t)$, given in Eq.~\eqref{eq:01}, the Schr\"odinger's equation results on two coupled equations
\ben
\frac{1}{2m} \left(\nabla S(q_i;t)\right)^{2}+ V(q_i;t) - \frac{\hbar^{2}}{2m}\frac{\nabla^{2}R(q_i;t)}{R(q_i;t)}&=&-\frac{\partial S(q_i;t)}{\partial t}\,, \nonumber \\
\label{eq:06}\\
\frac{\partial R^{2}(q_i;t)}{\partial t} + \nabla\cdot\left(R^{2}(q_i;t)\,\frac{\nabla S(q_i;t)}{m}\right)=0\,.
\label{eq:07}
\een
with $V(q_{i},t)$ being a external classical potential. Eqs. \eqref{eq:06} and \eqref{eq:07} describe the dynamic evolution of a particle in the classical theory and a continuity equation for the probability density, respectively, and the quantum nature of the events emerge from the coupled terms between these equations \cite{HOLLIVRO:93,bricmont2016broglie,sanz2019bohm}.

Eq.~\eqref{eq:06} provides a total energy, $-\frac{\partial S(q_i;t)}{\partial t}$, given by a sum of kinetic and potential  energies, plus an additional term interpreted as a quantum potential \cite{hasan2016effect,lentrodt2020ab,gonzalez2007quantum,gonzalez2009bohmian,gonzalez2004bohmian,gonzalez2009bohmian2,gonzalez2008effective}, while Eq.~\eqref{eq:07} can be identified as a continuity equation, with the probability density $R^{2}(q_i;t)$ and the current density given by
\be
{\bf J}=R^{2}(q_i;t)\,\frac{{\bf \nabla} S(q_i;t)}{m}\,.
\ee
 
The uniqueness of $\Psi(q_{i},t)$ is immediately verified in $R(q_i;t)$, for each pair $\lbrace q_{i},t\rbrace$; but not necessarily into $S(q_{i},t)$, since for each pair one can define a distinct set of these functions. However, if the functions $S(q_{i},t)$ differ from each other by integer multiples of $\hbar$, then the wave function $\Psi(q_i;t)$ will be unique, and the field $p_{i}$ defined as
\begin{eqnarray}
{p_{i}}= \nabla S(q_{i},t)
\label{eq:09}
\end{eqnarray}
shall has  uniqueness assured for each points $\lbrace q_{i},t\rbrace$.

In QTM, the Eqs. \eqref{eq:06} and \eqref{eq:07} control the dynamics of a system particles \cite{hasan2016effect,lentrodt2020ab,gonzalez2007quantum,gonzalez2009bohmian,gonzalez2004bohmian,gonzalez2009bohmian2,gonzalez2008effective}. In this scenario, the term 
\begin{eqnarray}
V(q_i;t) - \frac{\hbar^{2}}{2m}\frac{\nabla^{2}R\left(q_{i},t\right)}{R\left(q_{i},t\right)}
\label{eq:10}
\end{eqnarray}
provides an effective potential in which the particle is submitted. Therefore, the Eq. \eqref{eq:06} consists into the Hamilton-Jacobi equation \cite{dittrich2016hamilton}, unless a so-called quantum potential term 
\begin{eqnarray}
Q(q_{i},t)=-\frac{\hbar^{2}}{2m}\frac{\nabla^{2}R\left(q_{i},t\right)}{R\left(q_{i},t\right)}.
\label{eq:11}
\end{eqnarray}
This term arises from the interaction between the guiding wave $\Psi(q_{i},t)$ and the particle, and it is responsible for events of quantum nature during the  evolution of the physical system \cite{hasan2016effect,lentrodt2020ab,gonzalez2007quantum,gonzalez2009bohmian,gonzalez2004bohmian,gonzalez2009bohmian2,gonzalez2008effective}.

Since $R(q_{i},t)$, Eq.~\eqref{eq:02}, consists in a probability density, Eq. \eqref{eq:07} provides a continuity equation  associated to $R(q_{i},t)$. In this regard, the specification of $q_{i}(t)$ and the guiding wave $\Psi(q_{i},t)$, at a certain instant $t$, defines the state of an individual system. As demonstrated from Eq.~\eqref{eq:06}, $Q(q_{i},t)$ depends explicitly of $R(q_{i},t)$, and it is coupled with $S(q_{i},t)$ in such way that
\ben
\frac{\partial S (q_{i},t)}{\partial t}+\frac{1}{2m} \left(\nabla S (q_{i},t)\right)^{2}+ V(q_{i},t) + Q(q_{i},t) =0\,. \nonumber \\
\een
Thus, the quantum potential is not a previously known potential, such as $V(q_{i},t)$, but it depends on the state of the whole system, and it defines an interaction wave-particle that evolves according to the system dynamics which is mediated by a force like effect \cite{maddox2003estimating,becker2019asymmetry,keating2001force,batelaan2015dispersionless}. In this regard, the dynamic of the particle wavepacket can be described in terms of a effective force:
\begin{eqnarray}
\textbf{F}_{eff}=\frac{\mbox{d}{\bf{p}}}{\mbox{d}t}=\textbf{F}_{C}+\textbf{F}_{Q}~,
\label{eq:12a}
\end{eqnarray}
in terms of the classical force ($\textbf{F}_{C}$), derived from the classical potential $V(q_{i},t)$, and the so-called quantum force ($\textbf{F}_{Q}$) \cite{maddox2003estimating,becker2019asymmetry}
\begin{eqnarray}
\textbf{F}_Q(q_{i},t)=-\nabla Q(q_{i},t)~,
\label{eq:12}
\end{eqnarray}
derived from the quantum potential, Eq.~\eqref{eq:11}.

The quantum force acts on the de Broglie-Bohm trajectories \cite{kocsis2011observing}, and it is not mesurable \citep{kocsis2011observing,becker2019asymmetry}. In an operational way, the presence of the quantum force can be observed in the presence of a deflection in the average trajectories \cite{kocsis2011observing,becker2019asymmetry}. In this context, we propose a study of a free particle and a particle subjected to the Eckart potential, through the QTM, and so we compare the effect of a classical potential on the Bohm's quantum force.


\section{Temporal Propagation Through the Finite-Difference Method}\label{sec-2}

Most of the studies involving scattering in QTM searches for descriptive and representative quantities of the dynamic process~\cite{San525:12,keating2001force,gonzalez2007quantum,gonzalez2009bohmian,gonzalez2004bohmian}. These quantities are obtained in terms of the functions $R(q_{i},t)$ and $S(q_{i},t)$. Thus, one can  solve the  Schr\"odinger equation, and obtain these functions in terms of $\Psi(q_{i},t)$.  In this work, we apply the Quantum Trajectory Method~\cite{Wya,Wya187:99} on the field $\Psi$, in order to obtain the system dynamics through interactive processes at a given initial condition, with the proper adjustments to ensure the convergence criteria and stability. Additionaly, we have limited our applications in one-dimensional problems: the free particle and with the presence of a classical Eckart potential. 

Adopting the interactive finite-difference method \cite{simos1999finite,cooper2010finite,finite} {since it is a common method to the students, being widely discussed in traditional undergraduate courses in physics,} the one-dimensional time-dependent Schr\"odinger equation~ can be written as 
 
\begin{eqnarray}
\frac{\Psi(q,t+\Delta t)- \Psi(q,t)}{\Delta t}=\frac{i}{2}\frac{\partial^{2}\Psi(q,t)}{\partial q^{2}} - i V(q,t)\Psi(q,t), \nonumber \\
\label{eq:14}
\end{eqnarray}
where $\Delta t$ is a small finite time interval and we use the atomic units system, in order to ensure a reasonable performance without compromising the relevant theoretical aspects.

In order to make use of the propagation process, it is necessary to define the initial state of the quantum wave function. Here, we are choosing the Gaussian packet \cite{livre,belinfante1974survey} at the instant $t=0$, 
\begin{eqnarray}
\psi(q,0)=\left(\frac{2\gamma}{\pi}\right)^{\frac{1}{4}} \exp\left[-\gamma\left(q-q_{0}\right)^{2}+ip_{0}(q-q_{0})\right], \nonumber\\
\label{eq:15}
\end{eqnarray}
where $\gamma=1/2\delta^2$, with $\delta$ being the  packet's width, , and $q_0$ and $p_0$ are, respectively, the center of position and momentum of the packet.

Since the scalar fields $R(q,t)$ and $S(q,t)$ can be determined in terms of $\Psi(q, t)$, Eqs.~\eqref{eq:02} and \eqref{eq:03}, one may use them into Eqs. \eqref{eq:06} and \eqref{eq:07}, in order to obtain the dynamic of the system. Considering the problem under the influence of a time-independent potential $V(q)$ and the Eq.~\eqref{eq:09}, 
it is possible to determine the velocity distributions $\dot{q}(t)$ and the associated trajectories, as well as the effective force related to the quantum potential. For determination of the trajectory, we use the temporal propagation by finite differences technique, making the necessary adjustments for the initial conditions,
\begin{eqnarray}
q(t_{k}+\Delta t)&=&q(t_{k}) + \frac{\partial S(q,t_{k})}{\partial q}\,\Delta t \,.\nonumber 
\end{eqnarray}
In addition, for determination of the mediating force,  from Eq.~\eqref{eq:11} one can apply the finite differences approach upon the quantum potential $Q$ as
\begin{eqnarray}
Q(q, 0)&=& \frac{1}{2R(q,0)} \times\nonumber \\
&& \times \left[\frac{R(q+\Delta q,0)- 2R(q,0) + R(q-\Delta q,0)}{\Delta q ^{2}}\right] \nonumber \\\label{eq:11b}
\end{eqnarray}
 in terms of the generalized coordinates. In the cases considered in this work, the implementation of the numerical calculus 
with a discretization of $2 500$ points in the variable $q$ and  $10^{7}$  points in the variable $t$, in a way to guarantee a satisfactory description, without to incur significant divergences on the values, and assuring a relatively low computational cost \cite{finite}. 


\section{Results}\label{sec-3}
\label{sec-4}

\subsection{Free particle wavepacket}

\begin{figure}[h!]
\centering
{\includegraphics[scale=1.6]{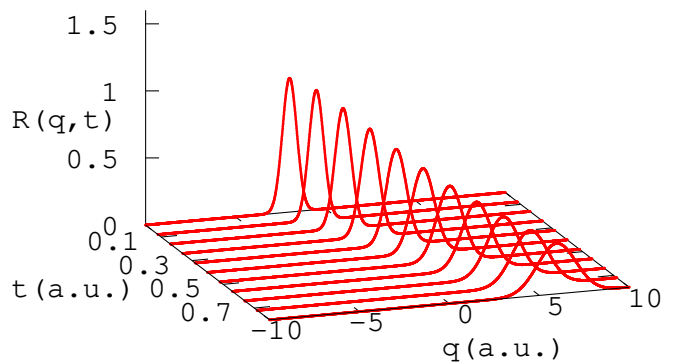}} \\
  \caption{(Color online) Propagation profile of a free Gaussian wavepacket, obained from the temporal propagation through the finite-difference method.} \label{Figura:01}
\end{figure}

Considering the propagation of a free ($V(q)=0 $) Gaussiam wavepacket \cite{livre}, Eq. (\ref{eq:15}),centerd on $q_{0}=-2.0\, a.u.$, $p_{0}=10\,a.u.$ and spatially distributed in the interval $[-10,10]$. We obtain the propagation profile for this wavepacket, applying the temporal propagation through the finite-difference method, (see Fig.~\ref{Figura:01}). According the Fig.~\ref{Figura:01}, the scattering effect on the wavepacket is clearly perceived during the process of temporal propagation. That outcome also is provided by usual interpretations of the quantum mechanics, and it is intrinsically connected to the uncertainty of observations in the Schr\"odinger representation for position.

\begin{figure}[h!]
\centering
{\includegraphics[scale=1.25]{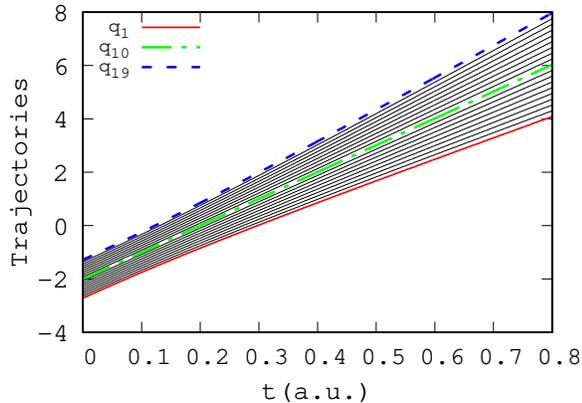}}
  \caption{(Color online) Trajectories associated to a set of nineteen points distributed over the free wavepacket, highlighting trajectories localized at the center $\lbrace q_{10} \rbrace$ (green dash-dotted line), left $\lbrace q_{1}\rbrace$ (solid red line) and rigth $q_{19} \rbrace$ (dashed blue line) of the packet.}\label{Figura:02}
\end{figure}

In order to highlight the trajectories localized at the center and extremes of the wavepacket we select nineteen points symmetrically distributed around the center of the wavepacket, $q_{0}=q_{10}$, which represents the initial configuration associated to the particles \textit{ensemble}. In such way, each point is initially distributed around $q_{10}$, as depicted in Fig.~\ref{Figura:02}.  Thus, through the dynamic variables we can observe what happen individually with the constituent elements of the distribution.

\begin{figure}[h!]
\centering
{\includegraphics[scale=1.3]{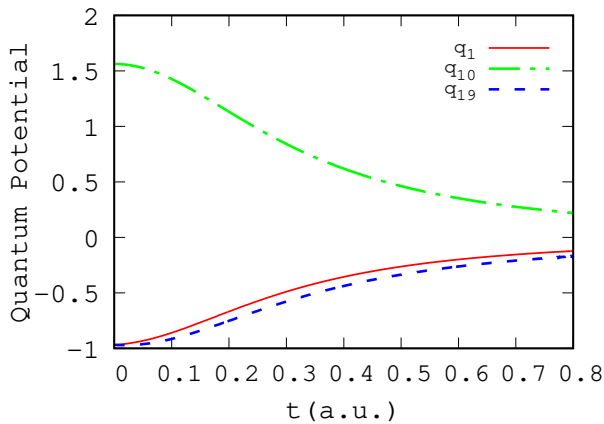}}
  \caption{(Color online) Quantum potential $Q(t)$ associated to the three representative trajectories: center (green dash-dotted line), extremes left (solid red line) and rigth (dashed blue line) of the free wavepacket.}\label{Figura:03}
\end{figure}

In the de Broglie-Bohm theory, despite the absence of a classic potential, the system is subjected to a quantum potential $Q(q,t)$, which arises from the dual wave-particle nature, through the interaction between the particle an its wave-guide. Thus, the  wavepacket propagation acquires a different connotation, which is explained as being a direct consequence of the action of a field $\Psi(q,t)$ on the \textit{ensemble} of particles via potential $Q(q,t)$, offering new prospects to the interpretation of the system dynamics. According to this representation, we calculate the quantum potential using Eq.~\eqref{eq:11b}. Fig.~\ref{Figura:03} shows the quantum potential associated to the three representative trajectories of the \textit{ensemble} at the center $\lbrace q_{10}\rbrace$ and extremes $\lbrace q_{1};q_{19} \rbrace$ of the free wavepacket. Those trajectories correspond to initial points localized at the center and extremes of the wavepacket, as highlighted in the Fig.~\ref{Figura:02}.

Therefore, from the quantum potential the \textit{ensemble} experiences the action of a non-null effective force (Eq.~\eqref{eq:12a}) consisting of elements intrinsically related to the initial conditions of the system, even in the absence of a classical potential, which evidences the non-classical nature of this process. Using Eq.~\eqref{eq:12} we calculate the respective quantum force associated to the same trajectories described in the Fig.~\ref{Figura:03}. In the absence of any classical potential, the effective force experienced by the wavepacket arises exclusively from the quantum potential being considered as a \textit{quantum force}. 

\begin{figure}[h]
\centering
\subfigure[]{\includegraphics[scale=1.3]{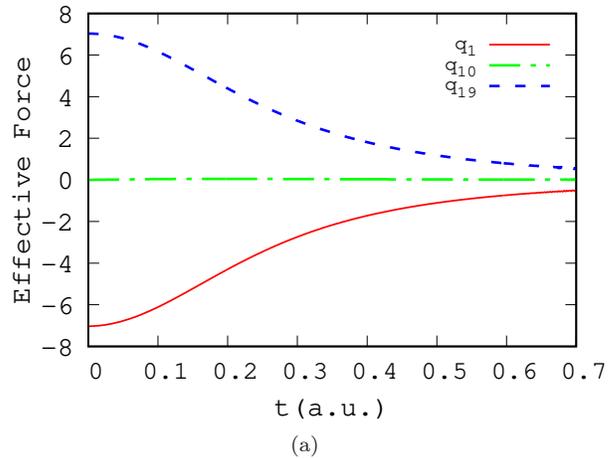}}
\subfigure[]{\includegraphics[scale=1.3]{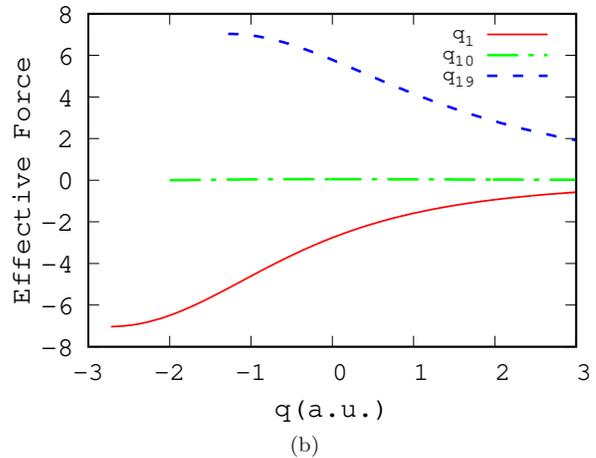}}
  \caption{(Color Online) Effective force as a function of the time (a) and the generalized coordinate $q(t)$ (b), for the trajectories localized at the center (green dash-dotted line), left (solid red line) and right (dashed blue line) of the free (gaussian) wavepacket. Since there is no classical potential the effective force is only due to the exitence of a quantum potential emergent from the interaction between the corpuscular and wave nature of the system.}\label{Figura:04}
\end{figure}

Fig.~\ref{Figura:04} shows the effective quantum force as a function of the time and the generalized coordinate $q(t)$. As indicated, although the effective force being zero at the center of wavepacket, the dispersion on the trajectories at the extremes obeys the tendency that the quantum force acts over the elements distributed at the edges of the wavepacket, in such way that it \textit{accelerates} \cite{newton} points on the left side of the wavepacket center (back), $q < q_{10}$, and \textit{slows down} \cite{newton} points on the right side (front), $q > q_{10}$. 
 
Therefore, one can conclude that the center of  the wavepacket experienced a classical free particle dynamics, because there is no classical or quantum force acting on it, whereas the edges experience a quantum dynamics from the quantum potential emergent from the interaction between the corpuscular and wave nature of the system. Therefore, the quantum force is strongly connected to the existence of the wavepacket itself, while the classical determinism of a physical system is in some way preserved, and the events of quantum nature are guided by a field of probabilistic nature, $\Psi$, which acts on the \textit{ensemble} of particle modifying the system dynamics, as a wave-guide.


\subsection{Particle subjected to the Eckart potential}

{In the perspective of to illustrate the effect of a classical potential on the quantum force \cite{gauss,dewdney1982quantum}, we consider the propagation of the wavepacket, Eq. (\ref{eq:15}), scattered by one of the most applicable and useful potentials for investigations about scattering parameters and bound states  \cite{gauss,razavy2003quantum,eckart1930penetration,johnston1962tunnelling,soylu2008kappa,Ikhdair_2014,Valencia_Ortega_2018,fern2019confluent,Mousavi_2019,dhali2019quantum}the Eckart potential}
\begin{eqnarray}
 V(q) = V_{0}\frac{\exp{\left[\beta (q-q_{v})\right]}}{\left\{1+\exp{\left[\beta (q-q_{v})\right]}\right\}^{2}},
 \label{eq:18}
\end{eqnarray}
where $V_0$, $\beta$ and $q_v$ are, respectively, amplitude, width and center of the potential.

Since, we have adopted atomic units, the coefficient $\beta$ has unit of inverse of the Bohr radius. For our analysis, we are assuming the potential with amplitude $V_{0}=200\,a.u.$, width $\beta = 20\,a.u.$, and centerd at $q_{0}=-2.0\,a.u.$ and $p_{0}=10\,a.u.$.

\begin{figure}[h]
\centering
\subfigure[]{\includegraphics[scale=1.5]{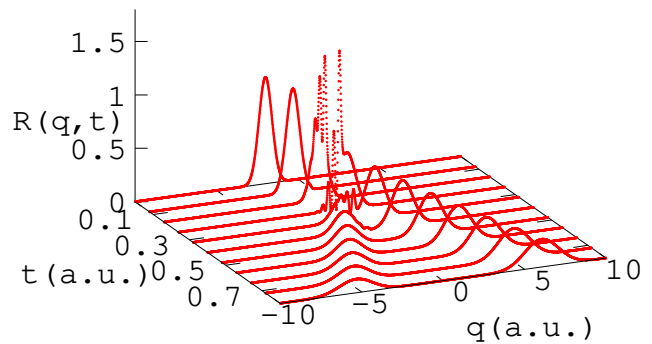}} \\
\subfigure[]{\includegraphics[scale=1.3]{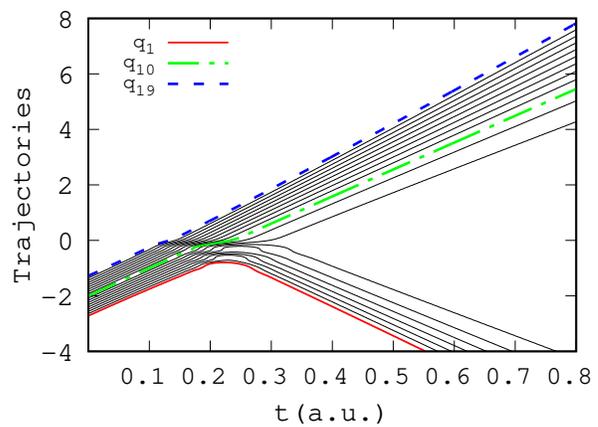}}
  \caption{(Color online) (a) Propagation profile of the Gaussian wavepacket subjected to the Eckart potential. (b) Trajectories of nineteen points distributed around $q_{0}=-2.0\,a.u.$, with $E=50.0\,a.u.$. The highlighted trajectories are associated to the points placed at the center (green dash-dotted line), extremes left (solid red line) and rigth (dashed blue line) of the wavepacket.}\label{Figura:06}
\end{figure}

In the Fig.~\ref{Figura:06} (a), we depicted the wavepacket propagation scattered by the Eckart potential given in the Eq.~\eqref{eq:18}, obtained from the finite-difference method. It furnishes the behavior characteristic for that type of process, showing a distinction for effects of transmission and reflection on this potential barrier. Even the initial average energy of the wavepacket being equal to the height of the barrier, one fraction of the packet is transmitted and the other one is reflected, with most of the amplitude being transmitted for present initial conditions. The propagation of the wavepacket and the dispersion, during the scattering process, are illustrated in terms of the trajectories pictured in the Fig.~\ref{Figura:06} (b).
Those trajectories, represented in Fig.~\ref{Figura:06} (b), allow us to conclude that the scattering process starts at $t=0.15\, a.u.$ and any influence  registered before this interval elapses without an explicit action of the classical potential. 
As illustrated, the left (back) of the wavepacket $\lbrace q_{1} \rbrace$ is reflected, whereas the center $\lbrace q_{10} \rbrace$ and the right (front) $\lbrace q_{19} \rbrace$ of the wavepacket are transmitted, tunneling the potential barrier. This effect can also be illustrated by the plot of the quantum potential and the analysis of the quantum forces acting in each trajectory.

\begin{figure}[h!]
\subfigure[]{\includegraphics[scale=1.3]{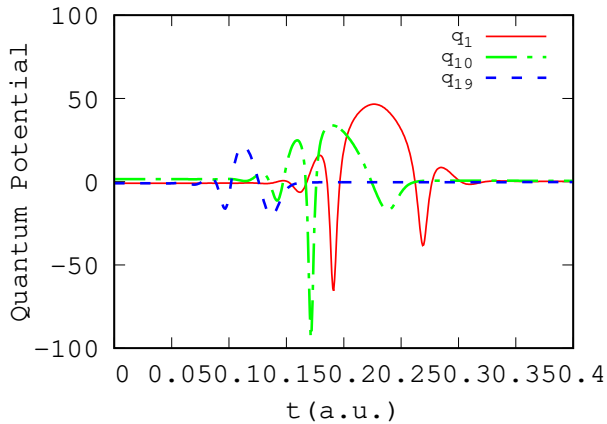}}
\subfigure[]{\includegraphics[scale=1.3]{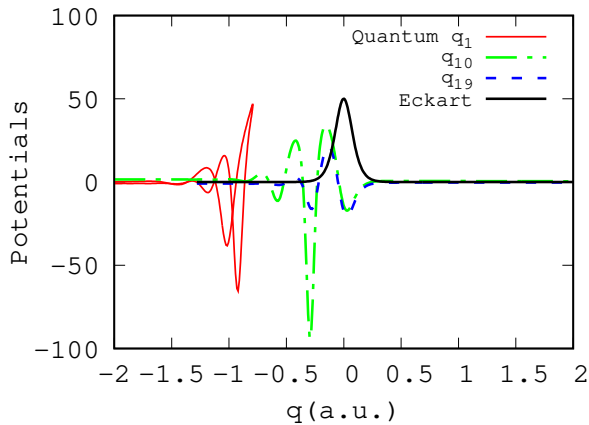}}
  \caption{(a) Quantum potential $Q(t)$ represented at the time interval $0\,a.u.$ and $0.15\,a.u.$, with $E=50.0\,a.u.$, subjected to the classical interaction of amplitude $V_0=200\, a.u.$. (b) Comparison between the quantum and classical potential. These profiles are associated to the three representative trajectories: center (green dash-dotted line), extremes left (solid red line) and rigth (dashed blue line) of the Gaussian wavepacket subjected to the Eckart potential.}\label{Figura:08}
\end{figure}

Fig.~\ref{Figura:08} shows the quantum potential $Q(q,t)$ obtained from  Eq.~\eqref{eq:11b} for the wavepacket scattered by a classical Eckart potential, for the trajectories highlighted in Fig.~\ref{Figura:06} (b), localized at the center $\lbrace q_{10} \rbrace$  and extremes $\lbrace q_{1};q_{19} \rbrace$ of the wavepacket. As shown in Fig.~\ref{Figura:08}, when the wavepacket approaches the potential  barrier the quantum potential profile changes and even before the scattering the behavior is completely different from the one obtained in the case of a free particle, considering the same quantities  (Fig.~\ref{Figura:03}). Fig.~\ref{Figura:08} (b) shows the tunneling of the front and the center of the wavepacket in the potential barrier as previously discussed. The tunnelling effect with the Eckart potential was discussed before in the literature in terms of the Bohmian Total Potential \cite{gonzalez2009bohmian,johnston1962tunnelling}.

\begin{figure}[h]
\centering
\subfigure[]{\includegraphics[scale=1.3]{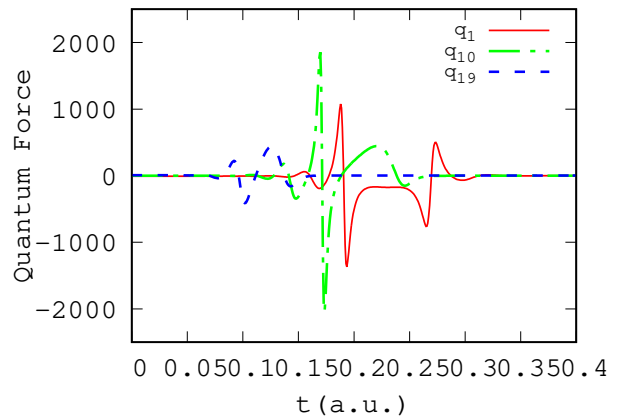}}
\subfigure[]{\includegraphics[scale=1.3]{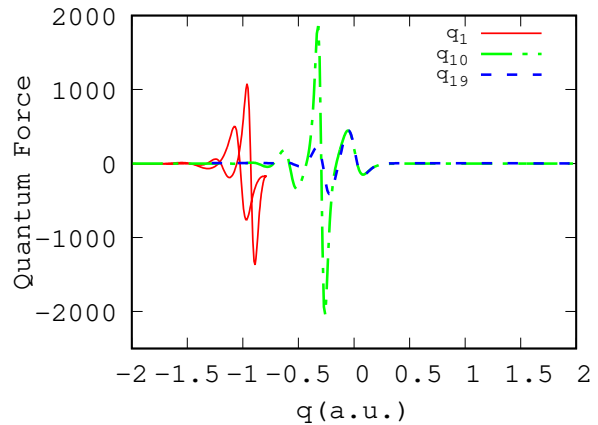}}
  \caption{(Color Online)(a) Quantum force as a function of the time and (b) the generalized coordinate $q(t)$, for the trajectories localized at the center (green dash-dotted line), left (solid red line) and rigth (dashed blue line) of the Gaussian wavepacket subjected to the Eckart potential.}\label{Figura:09}
\end{figure}

Using Eq.~\eqref{eq:12} we calculate the quantum force for the same trajectories described in Fig.~\ref{Figura:08}. Fig.~\ref{Figura:09} shows the quantum force as a function of the time and the generalized coordinate $q(t)$, for the wavepacket scattered by a classical Eckart potential. 

\begin{figure}[htp]
{\includegraphics[scale=1.3]{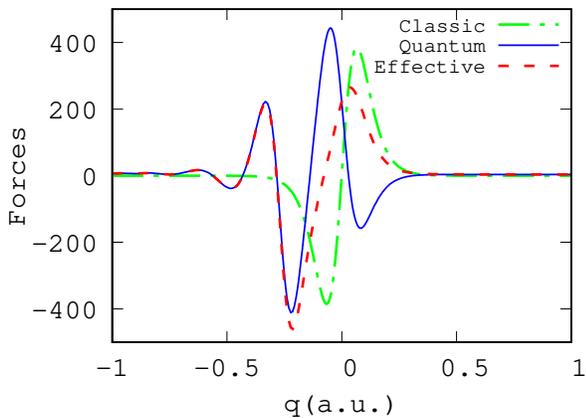}}
  \caption{(Color Online) Comparison between quantum (solid blue line), classic (green dash-dotted line) and effective force (dashed red line) for the trajectories localized at the rigth of the Gaussian wavepacket, subjected to the Eckart potential.}\label{Figura:10}
\end{figure}

As the scattering occurs, different points of the wavepacket experience a variation on the interaction profile which they are subjected. That can be seen in  Fig.~\ref{Figura:08} (b) describing the quantum potential $Q(q)$ and the classical potential $V(q)$ in terms of their coordinates. As can be seen, the constituents of the \textit{ensemble} perceive the classical potential  even before the classical interaction, since the element localized to the left side of the packet ($q_{1}$) suffers a significant change in its potential profile, even not interacting explicitly with the potential $V(q)$, but receiving this information through a correlation existing among the elements of the wavepacket. In other words, the particle experiences a quantum force effect which depends on the presence of the classical potential, even in the absence of any classical force field. In order to illustrate this effect we show in Fig.~\ref{Figura:10} comparison between the forces in the transmitted trajectory at the edge of the scattered wavepacket.

Therefore, this result can be interpreted analogously to that observed in the Aharonov-Bohm effect \cite{aharonov1959significance} since even in the absence of a force field, the quantum dynamics of the particle is altered by the presence of the classical potential. These results {strengthen the fact that classical potentials can act without force-fields, giving us indications that the Aharonov-Bohm effect could be observed in other classical potentials.}

\section{Conclusions}\label{sec-com}

{In this work, we report an application of the de Broglie-Bohm Quantum Theory of Motion as a powerful tool for evaluating Bohm's quantum force in the scattering process of a Gaussian wavepacket by a classical Eckart potential. In order to make our analysis easy to reproduce by undergraduate and graduate students of quantum mechanics courses, we adopt the temporal propagation method, which is an interactive technique of finite differences}. 

First, we consider the free particle dynamics, where we observe that, in the absence of a classical potential, the edges of the wavepacket experience an effective force effect, intrinsically related to the existence of the quantum potential $Q(q,t)$, which emerges from the interaction between the corpuscular and wave nature of the system, while the center of the wavepacket shows a classical free particle dynamics. Thus, the quantum force is strongly connected to the existence of the wavepacket itself, while the classical determinism of a physical system is in some way preserved. 

{In the following, we illustrate the effect of a classical Eckart potential, showing that the system experiences significant changes on its dynamics, even before the explicit interaction with the classical force, giving us evidences of the presence of the quantum force in the scattering process. Thus, the system experiences a quantum force effect, which depends on the classical potential, even in the absence of any classical force field, analogous to that observed in the Aharonov-Bohm effect, giving indications that the nature of this effect can be observed in different classical potentials.}

{Therefore, these results show the potential of the de Broglie-Bohm formulation as a complementary picture for the quantum theory, being a useful classroom working tool for study quantum dynamics through the concept of Bohm's quantum force, instead of merely an alternative interpretation of the quantum theory.}

\begin{acknowledgments}
W. S. Santana and F. V. Prudente would like to thank Mirco Ragni for the computational help. This study was financed in part by the CNPq and  the \textit{Coordena\c{c}\~{a}o de Aperfei\c{c}oamento de Pessoal de N\'{i}vel Superior - Brasil} (CAPES) - Finance Code 001.
\end{acknowledgments}


\begin{thebibliography}{47}%
\makeatletter
\providecommand \@ifxundefined [1]{%
 \@ifx{#1\undefined}
}%
\providecommand \@ifnum [1]{%
 \ifnum #1\expandafter \@firstoftwo
 \else \expandafter \@secondoftwo
 \fi
}%
\providecommand \@ifx [1]{%
 \ifx #1\expandafter \@firstoftwo
 \else \expandafter \@secondoftwo
 \fi
}%
\providecommand \natexlab [1]{#1}%
\providecommand \enquote  [1]{``#1''}%
\providecommand \bibnamefont  [1]{#1}%
\providecommand \bibfnamefont [1]{#1}%
\providecommand \citenamefont [1]{#1}%
\providecommand \href@noop [0]{\@secondoftwo}%
\providecommand \href [0]{\begingroup \@sanitize@url \@href}%
\providecommand \@href[1]{\@@startlink{#1}\@@href}%
\providecommand \@@href[1]{\endgroup#1\@@endlink}%
\providecommand \@sanitize@url [0]{\catcode `\\12\catcode `\$12\catcode
  `\&12\catcode `\#12\catcode `\^12\catcode `\_12\catcode `\%12\relax}%
\providecommand \@@startlink[1]{}%
\providecommand \@@endlink[0]{}%
\providecommand \url  [0]{\begingroup\@sanitize@url \@url }%
\providecommand \@url [1]{\endgroup\@href {#1}{\urlprefix }}%
\providecommand \urlprefix  [0]{URL }%
\providecommand \Eprint [0]{\href }%
\providecommand \doibase [0]{http://dx.doi.org/}%
\providecommand \selectlanguage [0]{\@gobble}%
\providecommand \bibinfo  [0]{\@secondoftwo}%
\providecommand \bibfield  [0]{\@secondoftwo}%
\providecommand \translation [1]{[#1]}%
\providecommand \BibitemOpen [0]{}%
\providecommand \bibitemStop [0]{}%
\providecommand \bibitemNoStop [0]{.\EOS\space}%
\providecommand \EOS [0]{\spacefactor3000\relax}%
\providecommand \BibitemShut  [1]{\csname bibitem#1\endcsname}%
\let\auto@bib@innerbib\@empty
\bibitem [{\citenamefont {Holland}(1993)}]{HOLLIVRO:93}%
  \BibitemOpen
  \bibfield  {author} {\bibinfo {author} {\bibfnamefont {P.~R.}\ \bibnamefont
  {Holland}},\ }\href@noop {} {\emph {\bibinfo {title} {The Quantum Theory of
  Motion: an account of de Broglie - Bohm causal interpretation of quantum
  mechanics}}}\ (\bibinfo  {publisher} {Cambridge University Press},\ \bibinfo
  {year} {1993})\BibitemShut {NoStop}%
\bibitem [{\citenamefont {Bricmont}(2016)}]{bricmont2016broglie}%
  \BibitemOpen
  \bibfield  {author} {\bibinfo {author} {\bibfnamefont {J.}~\bibnamefont
  {Bricmont}},\ }in\ \href@noop {} {\emph {\bibinfo {booktitle} {Making Sense
  of Quantum Mechanics}}}\ (\bibinfo  {publisher} {Springer},\ \bibinfo {year}
  {2016})\ pp.\ \bibinfo {pages} {129--197}\BibitemShut {NoStop}%
\bibitem [{\citenamefont {Sanz}(2019)}]{sanz2019bohm}%
  \BibitemOpen
  \bibfield  {author} {\bibinfo {author} {\bibfnamefont {A.}~\bibnamefont
  {Sanz}},\ }\href@noop {} {\bibfield  {journal} {\bibinfo  {journal}
  {Frontiers of Physics}\ }\textbf {\bibinfo {volume} {14}},\ \bibinfo {pages}
  {11301} (\bibinfo {year} {2019})}\BibitemShut {NoStop}%
\bibitem [{\citenamefont {Styer}\ \emph {et~al.}(2002)\citenamefont {Styer},
  \citenamefont {Balkin}, \citenamefont {Becker}, \citenamefont {Burns},
  \citenamefont {Dudley}, \citenamefont {Forth}, \citenamefont {Gaumer},
  \citenamefont {Kramer}, \citenamefont {Oertel}, \citenamefont {Park} \emph
  {et~al.}}]{styer2002nine}%
  \BibitemOpen
  \bibfield  {author} {\bibinfo {author} {\bibfnamefont {D.~F.}\ \bibnamefont
  {Styer}}, \bibinfo {author} {\bibfnamefont {M.~S.}\ \bibnamefont {Balkin}},
  \bibinfo {author} {\bibfnamefont {K.~M.}\ \bibnamefont {Becker}}, \bibinfo
  {author} {\bibfnamefont {M.~R.}\ \bibnamefont {Burns}}, \bibinfo {author}
  {\bibfnamefont {C.~E.}\ \bibnamefont {Dudley}}, \bibinfo {author}
  {\bibfnamefont {S.~T.}\ \bibnamefont {Forth}}, \bibinfo {author}
  {\bibfnamefont {J.~S.}\ \bibnamefont {Gaumer}}, \bibinfo {author}
  {\bibfnamefont {M.~A.}\ \bibnamefont {Kramer}}, \bibinfo {author}
  {\bibfnamefont {D.~C.}\ \bibnamefont {Oertel}}, \bibinfo {author}
  {\bibfnamefont {L.~H.}\ \bibnamefont {Park}},  \emph {et~al.},\ }\href@noop
  {} {\bibfield  {journal} {\bibinfo  {journal} {American Journal of Physics}\
  }\textbf {\bibinfo {volume} {70}},\ \bibinfo {pages} {288} (\bibinfo {year}
  {2002})}\BibitemShut {NoStop}%
\bibitem [{\citenamefont {Belinsky}(2019)}]{belinsky2019david}%
  \BibitemOpen
  \bibfield  {author} {\bibinfo {author} {\bibfnamefont {A.~V.}\ \bibnamefont
  {Belinsky}},\ }\href@noop {} {\bibfield  {journal} {\bibinfo  {journal}
  {Physics-Uspekhi}\ }\textbf {\bibinfo {volume} {62}},\ \bibinfo {pages}
  {1268} (\bibinfo {year} {2019})}\BibitemShut {NoStop}%
\bibitem [{\citenamefont {Bohm}(1952{\natexlab{a}})}]{bohm1952suggested}%
  \BibitemOpen
  \bibfield  {author} {\bibinfo {author} {\bibfnamefont {D.}~\bibnamefont
  {Bohm}},\ }\href@noop {} {\bibfield  {journal} {\bibinfo  {journal} {Physical
  review}\ }\textbf {\bibinfo {volume} {85}},\ \bibinfo {pages} {166} (\bibinfo
  {year} {1952}{\natexlab{a}})}\BibitemShut {NoStop}%
\bibitem [{\citenamefont {Bohm}(1952{\natexlab{b}})}]{bohm85suggested}%
  \BibitemOpen
  \bibfield  {author} {\bibinfo {author} {\bibfnamefont {D.}~\bibnamefont
  {Bohm}},\ }\href@noop {} {\bibfield  {journal} {\bibinfo  {journal} {Physical
  Review}\ }\textbf {\bibinfo {volume} {85}},\ \bibinfo {pages} {180} (\bibinfo
  {year} {1952}{\natexlab{b}})}\BibitemShut {NoStop}%
\bibitem [{\citenamefont {de~Broglie}()}]{de1928nouvelle}%
  \BibitemOpen
  \bibfield  {author} {\bibinfo {author} {\bibfnamefont {L.}~\bibnamefont
  {de~Broglie}},\ }\href@noop {} {\bibfield  {journal} {\bibinfo  {journal} {J.
  Bordet,(Gauthier-Villars, Paris, 1928)}\ ,\ \bibinfo {pages} {374}}}\bibinfo
  {note} {(English translation: Bacciagaluppi, G. and Valentini, A. (2009)
  \cite{bacciagaluppi2009quantum})}\BibitemShut {NoStop}%
\bibitem [{\citenamefont {Bacciagaluppi}\ and\ \citenamefont
  {Valentini}(2009)}]{bacciagaluppi2009quantum}%
  \BibitemOpen
  \bibfield  {author} {\bibinfo {author} {\bibfnamefont {G.}~\bibnamefont
  {Bacciagaluppi}}\ and\ \bibinfo {author} {\bibfnamefont {A.}~\bibnamefont
  {Valentini}},\ }\href@noop {} {\emph {\bibinfo {title} {Quantum theory at the
  crossroads: reconsidering the 1927 Solvay conference}}}\ (\bibinfo
  {publisher} {Cambridge University Press},\ \bibinfo {year} {2009})\ pp.\
  \bibinfo {pages} {30--88}\BibitemShut {NoStop}%
\bibitem [{\citenamefont {Pinto-Neto}\ and\ \citenamefont
  {Fabris}(2013)}]{pinto2013quantum}%
  \BibitemOpen
  \bibfield  {author} {\bibinfo {author} {\bibfnamefont {N.}~\bibnamefont
  {Pinto-Neto}}\ and\ \bibinfo {author} {\bibfnamefont {J.}~\bibnamefont
  {Fabris}},\ }\href@noop {} {\bibfield  {journal} {\bibinfo  {journal}
  {Classical and Quantum Gravity}\ }\textbf {\bibinfo {volume} {30}},\ \bibinfo
  {pages} {143001} (\bibinfo {year} {2013})}\BibitemShut {NoStop}%
\bibitem [{\citenamefont {Hasan}\ \emph {et~al.}(2016)\citenamefont {Hasan},
  \citenamefont {Hossen}, \citenamefont {Rafat},\ and\ \citenamefont
  {Mamun}}]{hasan2016effect}%
  \BibitemOpen
  \bibfield  {author} {\bibinfo {author} {\bibfnamefont {M.}~\bibnamefont
  {Hasan}}, \bibinfo {author} {\bibfnamefont {M.}~\bibnamefont {Hossen}},
  \bibinfo {author} {\bibfnamefont {A.}~\bibnamefont {Rafat}}, \ and\ \bibinfo
  {author} {\bibfnamefont {A.}~\bibnamefont {Mamun}},\ }\href@noop {}
  {\bibfield  {journal} {\bibinfo  {journal} {Chinese Physics B}\ }\textbf
  {\bibinfo {volume} {25}},\ \bibinfo {pages} {105203} (\bibinfo {year}
  {2016})}\BibitemShut {NoStop}%
\bibitem [{\citenamefont {Lentrodt}\ and\ \citenamefont
  {Evers}(2020)}]{lentrodt2020ab}%
  \BibitemOpen
  \bibfield  {author} {\bibinfo {author} {\bibfnamefont {D.}~\bibnamefont
  {Lentrodt}}\ and\ \bibinfo {author} {\bibfnamefont {J.}~\bibnamefont
  {Evers}},\ }\href@noop {} {\bibfield  {journal} {\bibinfo  {journal} {Phys.
  Rev. X}\ }\textbf {\bibinfo {volume} {10}},\ \bibinfo {pages} {011008}
  (\bibinfo {year} {2020})}\BibitemShut {NoStop}%
\bibitem [{\citenamefont {Gonz{\'a}lez}\ \emph {et~al.}(2007)\citenamefont
  {Gonz{\'a}lez}, \citenamefont {Gim{\'e}nez}, \citenamefont
  {Gonz{\'a}lez-Aguilar},\ and\ \citenamefont {Bofill}}]{gonzalez2007quantum}%
  \BibitemOpen
  \bibfield  {author} {\bibinfo {author} {\bibfnamefont {M.~F.}\ \bibnamefont
  {Gonz{\'a}lez}}, \bibinfo {author} {\bibfnamefont {X.}~\bibnamefont
  {Gim{\'e}nez}}, \bibinfo {author} {\bibfnamefont {J.}~\bibnamefont
  {Gonz{\'a}lez-Aguilar}}, \ and\ \bibinfo {author} {\bibfnamefont {J.~M.}\
  \bibnamefont {Bofill}},\ }\href@noop {} {\bibfield  {journal} {\bibinfo
  {journal} {J. Phys. Chem. A}\ }\textbf {\bibinfo {volume} {111}},\ \bibinfo
  {pages} {10226} (\bibinfo {year} {2007})}\BibitemShut {NoStop}%
\bibitem [{\citenamefont {Gonz{\'a}lez}\ \emph
  {et~al.}(2009{\natexlab{a}})\citenamefont {Gonz{\'a}lez}, \citenamefont
  {Bofill},\ and\ \citenamefont {Gim{\'e}nez}}]{gonzalez2009bohmian}%
  \BibitemOpen
  \bibfield  {author} {\bibinfo {author} {\bibfnamefont {M.~F.}\ \bibnamefont
  {Gonz{\'a}lez}}, \bibinfo {author} {\bibfnamefont {J.~M.}\ \bibnamefont
  {Bofill}}, \ and\ \bibinfo {author} {\bibfnamefont {X.}~\bibnamefont
  {Gim{\'e}nez}},\ }\href@noop {} {\bibfield  {journal} {\bibinfo  {journal}
  {J. Phys. Chem. A}\ }\textbf {\bibinfo {volume} {113}},\ \bibinfo {pages}
  {15024} (\bibinfo {year} {2009}{\natexlab{a}})}\BibitemShut {NoStop}%
\bibitem [{\citenamefont {Gonz{\'a}lez}\ \emph {et~al.}(2004)\citenamefont
  {Gonz{\'a}lez}, \citenamefont {Bofill},\ and\ \citenamefont
  {Gim{\'e}nez}}]{gonzalez2004bohmian}%
  \BibitemOpen
  \bibfield  {author} {\bibinfo {author} {\bibfnamefont {J.}~\bibnamefont
  {Gonz{\'a}lez}}, \bibinfo {author} {\bibfnamefont {J.~M.}\ \bibnamefont
  {Bofill}}, \ and\ \bibinfo {author} {\bibfnamefont {X.}~\bibnamefont
  {Gim{\'e}nez}},\ }\href@noop {} {\bibfield  {journal} {\bibinfo  {journal}
  {J. Chem. Phys.}\ }\textbf {\bibinfo {volume} {120}},\ \bibinfo {pages}
  {10961} (\bibinfo {year} {2004})}\BibitemShut {NoStop}%
\bibitem [{\citenamefont {Gonz{\'a}lez}\ \emph
  {et~al.}(2009{\natexlab{b}})\citenamefont {Gonz{\'a}lez}, \citenamefont
  {Aguilar-Mogas}, \citenamefont {Gonz{\'a}lez}, \citenamefont {Crehuet},
  \citenamefont {Anglada}, \citenamefont {Bofill},\ and\ \citenamefont
  {Gim{\'e}nez}}]{gonzalez2009bohmian2}%
  \BibitemOpen
  \bibfield  {author} {\bibinfo {author} {\bibfnamefont {M.~F.}\ \bibnamefont
  {Gonz{\'a}lez}}, \bibinfo {author} {\bibfnamefont {A.}~\bibnamefont
  {Aguilar-Mogas}}, \bibinfo {author} {\bibfnamefont {J.}~\bibnamefont
  {Gonz{\'a}lez}}, \bibinfo {author} {\bibfnamefont {R.}~\bibnamefont
  {Crehuet}}, \bibinfo {author} {\bibfnamefont {J.~M.}\ \bibnamefont
  {Anglada}}, \bibinfo {author} {\bibfnamefont {J.~M.}\ \bibnamefont {Bofill}},
  \ and\ \bibinfo {author} {\bibfnamefont {X.}~\bibnamefont {Gim{\'e}nez}},\
  }\href@noop {} {\bibfield  {journal} {\bibinfo  {journal} {Theor. Chem.
  Acc.}\ }\textbf {\bibinfo {volume} {123}},\ \bibinfo {pages} {51} (\bibinfo
  {year} {2009}{\natexlab{b}})}\BibitemShut {NoStop}%
\bibitem [{\citenamefont {Gonz{\'a}lez}\ \emph {et~al.}(2008)\citenamefont
  {Gonz{\'a}lez}, \citenamefont {Gim{\'e}nez}, \citenamefont {Gonz{\'a}lez},\
  and\ \citenamefont {Bofill}}]{gonzalez2008effective}%
  \BibitemOpen
  \bibfield  {author} {\bibinfo {author} {\bibfnamefont {M.~F.}\ \bibnamefont
  {Gonz{\'a}lez}}, \bibinfo {author} {\bibfnamefont {X.}~\bibnamefont
  {Gim{\'e}nez}}, \bibinfo {author} {\bibfnamefont {J.}~\bibnamefont
  {Gonz{\'a}lez}}, \ and\ \bibinfo {author} {\bibfnamefont {J.~M.}\
  \bibnamefont {Bofill}},\ }\href@noop {} {\bibfield  {journal} {\bibinfo
  {journal} {Journal of mathematical chemistry}\ }\textbf {\bibinfo {volume}
  {43}},\ \bibinfo {pages} {350} (\bibinfo {year} {2008})}\BibitemShut
  {NoStop}%
\bibitem [{\citenamefont {Becker}\ \emph {et~al.}(2019)\citenamefont {Becker},
  \citenamefont {Guzzinati}, \citenamefont {B{\'e}ch{\'e}}, \citenamefont
  {Verbeeck},\ and\ \citenamefont {Batelaan}}]{becker2019asymmetry}%
  \BibitemOpen
  \bibfield  {author} {\bibinfo {author} {\bibfnamefont {M.}~\bibnamefont
  {Becker}}, \bibinfo {author} {\bibfnamefont {G.}~\bibnamefont {Guzzinati}},
  \bibinfo {author} {\bibfnamefont {A.}~\bibnamefont {B{\'e}ch{\'e}}}, \bibinfo
  {author} {\bibfnamefont {J.}~\bibnamefont {Verbeeck}}, \ and\ \bibinfo
  {author} {\bibfnamefont {H.}~\bibnamefont {Batelaan}},\ }\href@noop {}
  {\bibfield  {journal} {\bibinfo  {journal} {Nature communications}\ }\textbf
  {\bibinfo {volume} {10}},\ \bibinfo {pages} {1} (\bibinfo {year}
  {2019})}\BibitemShut {NoStop}%
\bibitem [{\citenamefont {Sanz}(2015)}]{sanz2015investigating}%
  \BibitemOpen
  \bibfield  {author} {\bibinfo {author} {\bibfnamefont {A.}~\bibnamefont
  {Sanz}},\ }\href@noop {} {\bibfield  {journal} {\bibinfo  {journal}
  {Foundations of Physics}\ }\textbf {\bibinfo {volume} {45}},\ \bibinfo
  {pages} {1153} (\bibinfo {year} {2015})}\BibitemShut {NoStop}%
\bibitem [{\citenamefont {Batelaan}\ and\ \citenamefont
  {Becker}(2015)}]{batelaan2015dispersionless}%
  \BibitemOpen
  \bibfield  {author} {\bibinfo {author} {\bibfnamefont {H.}~\bibnamefont
  {Batelaan}}\ and\ \bibinfo {author} {\bibfnamefont {M.}~\bibnamefont
  {Becker}},\ }\href@noop {} {\bibfield  {journal} {\bibinfo  {journal} {EPL
  (Europhysics Letters)}\ }\textbf {\bibinfo {volume} {112}},\ \bibinfo {pages}
  {40006} (\bibinfo {year} {2015})}\BibitemShut {NoStop}%
\bibitem [{\citenamefont {Maddox}\ and\ \citenamefont
  {Bittner}(2003)}]{maddox2003estimating}%
  \BibitemOpen
  \bibfield  {author} {\bibinfo {author} {\bibfnamefont {J.~B.}\ \bibnamefont
  {Maddox}}\ and\ \bibinfo {author} {\bibfnamefont {E.~R.}\ \bibnamefont
  {Bittner}},\ }\href@noop {} {\bibfield  {journal} {\bibinfo  {journal} {J.
  Chem. Phys.}\ }\textbf {\bibinfo {volume} {119}},\ \bibinfo {pages} {6465}
  (\bibinfo {year} {2003})}\BibitemShut {NoStop}%
\bibitem [{\citenamefont {Shelankov}(1998)}]{shelankov1998magnetic}%
  \BibitemOpen
  \bibfield  {author} {\bibinfo {author} {\bibfnamefont {A.}~\bibnamefont
  {Shelankov}},\ }\href@noop {} {\bibfield  {journal} {\bibinfo  {journal} {EPL
  (Europhysics Letters)}\ }\textbf {\bibinfo {volume} {43}},\ \bibinfo {pages}
  {623} (\bibinfo {year} {1998})}\BibitemShut {NoStop}%
\bibitem [{\citenamefont {Berry}(1999)}]{berry1999aharonov}%
  \BibitemOpen
  \bibfield  {author} {\bibinfo {author} {\bibfnamefont {M.}~\bibnamefont
  {Berry}},\ }\href@noop {} {\bibfield  {journal} {\bibinfo  {journal} {J.
  Phys. A: Math. Gen.}\ }\textbf {\bibinfo {volume} {32}},\ \bibinfo {pages}
  {5627} (\bibinfo {year} {1999})}\BibitemShut {NoStop}%
\bibitem [{\citenamefont {Keating}\ and\ \citenamefont
  {Robbins}(2001)}]{keating2001force}%
  \BibitemOpen
  \bibfield  {author} {\bibinfo {author} {\bibfnamefont {J.}~\bibnamefont
  {Keating}}\ and\ \bibinfo {author} {\bibfnamefont {J.}~\bibnamefont
  {Robbins}},\ }\href@noop {} {\bibfield  {journal} {\bibinfo  {journal} {J.
  Phys. A: Math. Gen.}\ }\textbf {\bibinfo {volume} {34}},\ \bibinfo {pages}
  {807} (\bibinfo {year} {2001})}\BibitemShut {NoStop}%
\bibitem [{\citenamefont {Aharonov}\ and\ \citenamefont
  {Bohm}(1959)}]{aharonov1959significance}%
  \BibitemOpen
  \bibfield  {author} {\bibinfo {author} {\bibfnamefont {Y.}~\bibnamefont
  {Aharonov}}\ and\ \bibinfo {author} {\bibfnamefont {D.}~\bibnamefont
  {Bohm}},\ }\href@noop {} {\bibfield  {journal} {\bibinfo  {journal} {Phys.
  Rev.}\ }\textbf {\bibinfo {volume} {115}},\ \bibinfo {pages} {485} (\bibinfo
  {year} {1959})}\BibitemShut {NoStop}%
\bibitem [{liv()}]{livre}%
  \BibitemOpen
  \href@noop {} {}\bibinfo {note} {For a detailed account, see Appendix G of
  Ref. \cite{belinfante1974survey}, this text gives analytical solutions for
  the Gaussian wavepacket describing the diffrences between the classical and
  the de Broglie-Bohm trajectories.}\BibitemShut {Stop}%
\bibitem [{\citenamefont {Belinfante}\ and\ \citenamefont
  {Ballentine}(1974)}]{belinfante1974survey}%
  \BibitemOpen
  \bibfield  {author} {\bibinfo {author} {\bibfnamefont {F.~J.}\ \bibnamefont
  {Belinfante}}\ and\ \bibinfo {author} {\bibfnamefont {L.~E.}\ \bibnamefont
  {Ballentine}},\ }\href@noop {} {\bibfield  {journal} {\bibinfo  {journal}
  {PhT}\ }\textbf {\bibinfo {volume} {27}},\ \bibinfo {pages} {53} (\bibinfo
  {year} {1974})}\BibitemShut {NoStop}%
\bibitem [{gau()}]{gauss}%
  \BibitemOpen
  \href@noop {} {}\bibinfo {note} {The scattering process of Gaussian
  wavepackets from classical potentials has been extensively studied using the
  de Broglie-Bohm theory over many years in the specialized literature that
  discusses quantum tunneling through Bohm's theory
  \cite{gonzalez2009bohmian,johnston1962tunnelling,razavy2003quantum}. For a
  detailed account, see ref. \cite{dewdney1982quantum}.}\BibitemShut {Stop}%
\bibitem [{\citenamefont {Dewdney}\ and\ \citenamefont
  {Hiley}(1982)}]{dewdney1982quantum}%
  \BibitemOpen
  \bibfield  {author} {\bibinfo {author} {\bibfnamefont {C.}~\bibnamefont
  {Dewdney}}\ and\ \bibinfo {author} {\bibfnamefont {B.~J.}\ \bibnamefont
  {Hiley}},\ }\href@noop {} {\bibfield  {journal} {\bibinfo  {journal}
  {Foundations of Physics}\ }\textbf {\bibinfo {volume} {12}},\ \bibinfo
  {pages} {27} (\bibinfo {year} {1982})}\BibitemShut {NoStop}%
\bibitem [{fin()}]{finite}%
  \BibitemOpen
  \href@noop {} {}\bibinfo {note} {The template of the program is not provided,
  since the prior programming experience of the readers could vary, and
  finite-difference method can be easily implemented in the most popular
  programming platforms as C++, FORTRAN, Python (using NumPy), MATLAB and
  Mathematica, for instance.}\BibitemShut {Stop}%
\bibitem [{\citenamefont {Dittrich}\ and\ \citenamefont
  {Reuter}(2016)}]{dittrich2016hamilton}%
  \BibitemOpen
  \bibfield  {author} {\bibinfo {author} {\bibfnamefont {W.}~\bibnamefont
  {Dittrich}}\ and\ \bibinfo {author} {\bibfnamefont {M.}~\bibnamefont
  {Reuter}},\ }in\ \href@noop {} {\emph {\bibinfo {booktitle} {Classical and
  Quantum Dynamics}}}\ (\bibinfo  {publisher} {Springer},\ \bibinfo {year}
  {2016})\ pp.\ \bibinfo {pages} {75--92}\BibitemShut {NoStop}%
\bibitem [{\citenamefont {Kocsis}\ \emph {et~al.}(2011)\citenamefont {Kocsis},
  \citenamefont {Braverman}, \citenamefont {Ravets}, \citenamefont {Stevens},
  \citenamefont {Mirin}, \citenamefont {Shalm},\ and\ \citenamefont
  {Steinberg}}]{kocsis2011observing}%
  \BibitemOpen
  \bibfield  {author} {\bibinfo {author} {\bibfnamefont {S.}~\bibnamefont
  {Kocsis}}, \bibinfo {author} {\bibfnamefont {B.}~\bibnamefont {Braverman}},
  \bibinfo {author} {\bibfnamefont {S.}~\bibnamefont {Ravets}}, \bibinfo
  {author} {\bibfnamefont {M.~J.}\ \bibnamefont {Stevens}}, \bibinfo {author}
  {\bibfnamefont {R.~P.}\ \bibnamefont {Mirin}}, \bibinfo {author}
  {\bibfnamefont {L.~K.}\ \bibnamefont {Shalm}}, \ and\ \bibinfo {author}
  {\bibfnamefont {A.~M.}\ \bibnamefont {Steinberg}},\ }\href@noop {} {\bibfield
   {journal} {\bibinfo  {journal} {Science}\ }\textbf {\bibinfo {volume}
  {332}},\ \bibinfo {pages} {1170} (\bibinfo {year} {2011})}\BibitemShut
  {NoStop}%
\bibitem [{\citenamefont {Sanz}\ and\ \citenamefont
  {Miret$-$Art\'{e}s}(2012)}]{San525:12}%
  \BibitemOpen
  \bibfield  {author} {\bibinfo {author} {\bibfnamefont {A.}~\bibnamefont
  {Sanz}}\ and\ \bibinfo {author} {\bibfnamefont {S.}~\bibnamefont
  {Miret$-$Art\'{e}s}},\ }\href@noop {} {\bibfield  {journal} {\bibinfo
  {journal} {Am. J. Phys.}\ }\textbf {\bibinfo {volume} {80}},\ \bibinfo
  {pages} {525} (\bibinfo {year} {2012})}\BibitemShut {NoStop}%
\bibitem [{\citenamefont {Wyatt}(1999{\natexlab{a}})}]{Wya}%
  \BibitemOpen
  \bibfield  {author} {\bibinfo {author} {\bibfnamefont {R.~E.}\ \bibnamefont
  {Wyatt}},\ }\href@noop {} {\bibfield  {journal} {\bibinfo  {journal} {J.
  Chem. Phys.}\ }\textbf {\bibinfo {volume} {111}},\ \bibinfo {pages} {4406}
  (\bibinfo {year} {1999}{\natexlab{a}})}\BibitemShut {NoStop}%
\bibitem [{\citenamefont {Wyatt}(1999{\natexlab{b}})}]{Wya187:99}%
  \BibitemOpen
  \bibfield  {author} {\bibinfo {author} {\bibfnamefont {R.~E.}\ \bibnamefont
  {Wyatt}},\ }\href@noop {} {\bibfield  {journal} {\bibinfo  {journal} {Chem.
  Phys. Lett.}\ }\textbf {\bibinfo {volume} {313}},\ \bibinfo {pages} {189}
  (\bibinfo {year} {1999}{\natexlab{b}})}\BibitemShut {NoStop}%
\bibitem [{\citenamefont {Simos}\ and\ \citenamefont
  {Williams}(1999)}]{simos1999finite}%
  \BibitemOpen
  \bibfield  {author} {\bibinfo {author} {\bibfnamefont {T.~E.}\ \bibnamefont
  {Simos}}\ and\ \bibinfo {author} {\bibfnamefont {P.~S.}\ \bibnamefont
  {Williams}},\ }\href@noop {} {\bibfield  {journal} {\bibinfo  {journal}
  {Computers \& chemistry}\ }\textbf {\bibinfo {volume} {23}},\ \bibinfo
  {pages} {513} (\bibinfo {year} {1999})}\BibitemShut {NoStop}%
\bibitem [{\citenamefont {Cooper}\ \emph {et~al.}(2010)\citenamefont {Cooper},
  \citenamefont {Valavanis}, \citenamefont {Ikoni{\'c}}, \citenamefont
  {Harrison},\ and\ \citenamefont {Cunningham}}]{cooper2010finite}%
  \BibitemOpen
  \bibfield  {author} {\bibinfo {author} {\bibfnamefont {J.}~\bibnamefont
  {Cooper}}, \bibinfo {author} {\bibfnamefont {A.}~\bibnamefont {Valavanis}},
  \bibinfo {author} {\bibfnamefont {Z.}~\bibnamefont {Ikoni{\'c}}}, \bibinfo
  {author} {\bibfnamefont {P.}~\bibnamefont {Harrison}}, \ and\ \bibinfo
  {author} {\bibfnamefont {J.}~\bibnamefont {Cunningham}},\ }\href@noop {}
  {\bibfield  {journal} {\bibinfo  {journal} {J. Appl. Phys.}\ }\textbf
  {\bibinfo {volume} {108}},\ \bibinfo {pages} {113109} (\bibinfo {year}
  {2010})}\BibitemShut {NoStop}%
\bibitem [{new()}]{newton}%
  \BibitemOpen
  \href@noop {} {}\bibinfo {note} {Although we have adopted the idea of an
  acceleration, the effective force does not consist of a Newtonian interaction
  between the wave-guide and the particle, so it is not possible to talk about
  a pair of action and reaction between them.}\BibitemShut {Stop}%
\bibitem [{\citenamefont {Razavy}(2003)}]{razavy2003quantum}%
  \BibitemOpen
  \bibfield  {author} {\bibinfo {author} {\bibfnamefont {M.}~\bibnamefont
  {Razavy}},\ }\href@noop {} {\emph {\bibinfo {title} {Quantum theory of
  tunneling}}}\ (\bibinfo  {publisher} {World Scientific},\ \bibinfo {year}
  {2003})\BibitemShut {NoStop}%
\bibitem [{\citenamefont {Eckart}(1930)}]{eckart1930penetration}%
  \BibitemOpen
  \bibfield  {author} {\bibinfo {author} {\bibfnamefont {C.}~\bibnamefont
  {Eckart}},\ }\href@noop {} {\bibfield  {journal} {\bibinfo  {journal} {Phys.
  Rev.}\ }\textbf {\bibinfo {volume} {35}},\ \bibinfo {pages} {1303} (\bibinfo
  {year} {1930})}\BibitemShut {NoStop}%
\bibitem [{\citenamefont {Johnston}\ and\ \citenamefont
  {Heicklen}(1962)}]{johnston1962tunnelling}%
  \BibitemOpen
  \bibfield  {author} {\bibinfo {author} {\bibfnamefont {H.~S.}\ \bibnamefont
  {Johnston}}\ and\ \bibinfo {author} {\bibfnamefont {J.}~\bibnamefont
  {Heicklen}},\ }\href@noop {} {\bibfield  {journal} {\bibinfo  {journal} {J.
  Phys. Chem.}\ }\textbf {\bibinfo {volume} {66}},\ \bibinfo {pages} {532}
  (\bibinfo {year} {1962})}\BibitemShut {NoStop}%
\bibitem [{\citenamefont {Soylu}\ \emph {et~al.}(2008)\citenamefont {Soylu},
  \citenamefont {Bayrak},\ and\ \citenamefont {Boztosun}}]{soylu2008kappa}%
  \BibitemOpen
  \bibfield  {author} {\bibinfo {author} {\bibfnamefont {A.}~\bibnamefont
  {Soylu}}, \bibinfo {author} {\bibfnamefont {O.}~\bibnamefont {Bayrak}}, \
  and\ \bibinfo {author} {\bibfnamefont {I.}~\bibnamefont {Boztosun}},\
  }\href@noop {} {\bibfield  {journal} {\bibinfo  {journal} {Journal of Physics
  A: Mathematical and Theoretical}\ }\textbf {\bibinfo {volume} {41}},\
  \bibinfo {pages} {065308} (\bibinfo {year} {2008})}\BibitemShut {NoStop}%
\bibitem [{\citenamefont {Ikhdair}\ and\ \citenamefont
  {Falaye}(2014)}]{Ikhdair_2014}%
  \BibitemOpen
  \bibfield  {author} {\bibinfo {author} {\bibfnamefont {S.~M.}\ \bibnamefont
  {Ikhdair}}\ and\ \bibinfo {author} {\bibfnamefont {B.~J.}\ \bibnamefont
  {Falaye}},\ }\href@noop {} {\bibfield  {journal} {\bibinfo  {journal} {Eur.
  Phys. J. Plus}\ }\textbf {\bibinfo {volume} {129}} (\bibinfo {year}
  {2014})}\BibitemShut {NoStop}%
\bibitem [{\citenamefont {Valencia-Ortega}\ and\ \citenamefont
  {Arias-Hernandez}(2018)}]{Valencia_Ortega_2018}%
  \BibitemOpen
  \bibfield  {author} {\bibinfo {author} {\bibfnamefont {G.}~\bibnamefont
  {Valencia-Ortega}}\ and\ \bibinfo {author} {\bibfnamefont {L.-A.}\
  \bibnamefont {Arias-Hernandez}},\ }\href@noop {} {\bibfield  {journal}
  {\bibinfo  {journal} {Int. J. Quantum Chem.}\ }\textbf {\bibinfo {volume}
  {118}},\ \bibinfo {pages} {e25589} (\bibinfo {year} {2018})}\BibitemShut
  {NoStop}%
\bibitem [{\citenamefont {Roy}\ and\ \citenamefont
  {C.}(2020)}]{fern2019confluent}%
  \BibitemOpen
  \bibfield  {author} {\bibinfo {author} {\bibfnamefont {B.}~\bibnamefont
  {Roy}}\ and\ \bibinfo {author} {\bibfnamefont {D.~J.~F.}\ \bibnamefont
  {C.}},\ }\href@noop {} {\bibfield  {journal} {\bibinfo  {journal} {Physica
  Scripta}\ }\textbf {\bibinfo {volume} {95}},\ \bibinfo {pages} {055210}
  (\bibinfo {year} {2020})}\BibitemShut {NoStop}%
\bibitem [{\citenamefont {Mousavi}\ and\ \citenamefont
  {Shojaei}(2019)}]{Mousavi_2019}%
  \BibitemOpen
  \bibfield  {author} {\bibinfo {author} {\bibfnamefont {M.}~\bibnamefont
  {Mousavi}}\ and\ \bibinfo {author} {\bibfnamefont {M.~R.}\ \bibnamefont
  {Shojaei}},\ }\href@noop {} {\bibfield  {journal} {\bibinfo  {journal}
  {Modern Physics Letters A}\ }\textbf {\bibinfo {volume} {34}},\ \bibinfo
  {pages} {1950073} (\bibinfo {year} {2019})}\BibitemShut {NoStop}%
\bibitem [{\citenamefont {Dhali}\ \emph {et~al.}(2019)\citenamefont {Dhali},
  \citenamefont {John},\ and\ \citenamefont {Swathi}}]{dhali2019quantum}%
  \BibitemOpen
  \bibfield  {author} {\bibinfo {author} {\bibfnamefont {R.}~\bibnamefont
  {Dhali}}, \bibinfo {author} {\bibfnamefont {C.}~\bibnamefont {John}}, \ and\
  \bibinfo {author} {\bibfnamefont {R.~S.}\ \bibnamefont {Swathi}},\
  }\href@noop {} {\bibfield  {journal} {\bibinfo  {journal} {J. Phys. Chem. A}\
  }\textbf {\bibinfo {volume} {123}},\ \bibinfo {pages} {7499} (\bibinfo {year}
  {2019})}\BibitemShut {NoStop}%
\end{thebibliography}
\end{document}